\documentclass[aps,pra,twocolumn,showpacs]{revtex4-2}

\usepackage{color,natbib}
\newcommand{\tcr}[1]{\textcolor{black}{#1}}

\usepackage{graphicx,amsmath,amssymb}

\begin{document}

\title{Certification \tcr{of stellar} ranks of quantum states of light\\ with a pair of click detectors}

\author{Jaromír Fiurášek}
\affiliation{Department of Optics, Faculty of Science, Palack\'y University, 17.\ listopadu 12, 77900  Olomouc, Czech Republic}

\begin{abstract}
Stellar rank of quantum state of light quantifes the amount of non-Gaussian resources required for its generation. One popular and practical approach to certification of stellar rank is based on measurement of click statistics with an array of binary detectors that can only distinguish the presence and absence of photons.  Specifically, it was shown that measurements with an array of $m+1$ detectors allow one to certify stellar rank $m$ of approximate Fock state $|m\rangle$, even when the state is subjected to losses or certain noise. 
Here we address the question as to how many click detectors are in principle required to certify stellar ranks \tcr{higher than one}.  
We show that two click detectors arranged in a  Hanbury Brown-Twiss setup suffice. Interestingly, detection of stellar ranks \tcr{higher than one} is greatly facilitated by making the total detection efficiency of the detectors sufficiently low but well calibrated.
Losses affect the response of the detection scheme in a way that can be exploited to certify stellar rank of higher Fock states. We explicitly construct the corresponding stellar rank witnesses and  discuss dependence of the stellar rank thresholds on parameters of the considered setup. Our results reveal that it is possible to certify stellar ranks higher than $1$ even with a  minimalistic scheme that provides only very  coarse-grained information about the photon number statistics of the characterized state.

\end{abstract}

\maketitle

\section{Introduction}

Generation of highly non-classical states of light such as Fock states \cite{Lvovsky2001,Cooper2013,Bouillard2019,Kawasaki2022}, various catlike states \cite{Ourjoumtsev2006,Nielsen2006,Ourjoumtsev2007,Wakui2007,Takahashi2008,Huang2015,Sychev2017,Etesse2015,Cotte2022,Simon2024,Endo2025}, approximate Gottesman-Kitaev-Preskill states \cite{Konno2024,Xanadu2025}, states exhibiting nonlinear squeezing \cite{Konno2021,Kala2025}, or  other interesting states \cite{Nielsen2010,Yukawa2013,Costanzo2017} requires non-Gaussian operations  \cite{Lvovsky2020,Walschaers2021,Biagi2022} in addition to Gaussian squeezing,  coherent displacements, passive linear optics, and homodyne detection \cite{Braunstein2005,Weedbrook2012}. The non-Gaussian resources required for preparation of a given quantum state of light can be quantified by its stellar rank \cite{Chabaud2020,Chabaud2021}.  Finite superpositions of Fock states possess stellar rank equal to the number of the  highest Fock state in the superposition. Stellar rank is invariant under Gaussian unitary operations and Gaussian states have stellar rank equal to zero. There exist states with infinite stellar rank such as superpositions of coherent states. The concept of stellar rank is equivalent to the genuine $n$-photon quantum non-Gaussianity introduced in Ref. \cite{Lachman2019}.

In practical terms, the stellar rank quantifies the number of elementary photon additions \cite{Zavatta2004,Barbieri2010,Kumar2013,Fadrny2024,Chen2024} or photon subtractions \cite{Ourjoumtsev2006,Nielsen2006,Wakui2007} that have to be performed to generate the state, following, e.g., the protocols proposed in Refs. \cite{Dakna1999,Fiurasek2005}. Equivalently, the stellar rank defines the minimum number of photons that have to be detected when the state is prepared by photon counting measurements on a part of multimode Gaussian entangled state \cite{Motamedi2025}. The simplest yet illustrative example is the conditional preparation of Fock state $|n\rangle$  by detection of $n$ photons in the idler mode of two-mode squeezed vacuum state \cite{Lvovsky2001,Cooper2013,Bouillard2019,Kawasaki2022}.

Stellar rank can be certified by suitable stellar rank witnesses \cite{Lachman2019,Chabaud2021,Fiurasek2022}. Various witnesses of stellar rank have been proposed in the literature and tested experimentally \cite{Chabaud2021,Fiurasek2022,Lachman2019,Podhora2022,Fadel2025,Lachman2025,Laurat2025,Kovalenko2024,Provaznik2025}. The stellar rank witnesses represent generalization and extension of witnesses of quantum non-Gaussianity 
\cite{Filip2011,Genoni2013,Lachman2013,Hughes2014,Lachman2016,Park2017,Happ2018,Kuhn2018,Fiurasek2021,Lachman2022,Marek2024,Brauer2025a,Brauer2025b,Kala2025b}. Those latter witnesses can certify that the state is not a Gaussian state or a mixture of Gaussian states and thus possesses nonzero stellar rank.

Quantum states of light are often detected with binary click detectors that can only distinguish the presence or absence of photons. To enable approximate photon counting one can utilize spatial or temporal multiplexing \cite{Paul1996,Rehacek2003,Banaszek2003,Fitch2003,Achilles2003,Bartley2013,Kalashnikov2011,Sperling2012a,Sperling2012b,Mattioli2016,Kroger2017,Sperling2017b,%
Zhu2018,Straka2018,Tiedau2019,Lachman2019,Hlousek2019,Cheng2022,Hlousek2024,Sullivan2024,Krishnaswamy2024,Santana2024} and split the input signal onto an array of $M$  detectors and measure the probabilities that a particular set of $k$ detectors click. In  Ref. \cite{Lachman2019}, stellar rank witnesses based on linear combinations of clicks of $m$ detectors and clicks of $m+1$ detectors for array of $m+1$ click detectors were constructed and utilized to certify the stellar rank $m$ (or equivalently the genuine $m$-photon quantum non-Gaussianity) of multipohoton Fock states.

Certification of stellar rank $m$ of Fock state $|m\rangle$  by coincidence measurements with $M>m$ binary click detectors is a natural choice since such detection reasonably well approximates the ideal photon counting. One can nevertheless wonder whether it would be possible to certify stellar rank of Fock states even with simpler experimental configuration and smaller number of binary  detectors. Here we answer this question in affirmative and we show that it suffices to employ a scheme with two binary detectors. 
Interestingly, detection of stellar ranks higher than one is greatly facilitated by making the total detection efficiency sufficiently small which can be ensured by inserting additional losses in front of the detectors. It has been proposed in the past that photon number distribution of a quantum state can be reconstructed from measurements with a single binary click detector preceded with tunable losses \cite{Mogilevtsev1998,Rossi2004}. By contrast, here the losses are fixed and a single measurement configuration is utilized to certify the stellar rank. While the losses are helpful, they are not strictly necessary and we show that an unbalanced detection scheme can certify stellar rank  higher than one even with perfect detectors. However, even in this case losses can make the detection of higher stellar ranks more experimentally feasible.

\tcr{Recently, quantum non-Gaussianity criteria (i.e., criteria for certification of stellar rank one) that account for losses of click detectors were presented in Ref. \cite{Checchinato2024} and utilized to certify quantum non-Gaussianity of photons emitted from quantum dot. Inclusion of detection efficiency into the description of the detector makes the resulting quantum non-Gaussianity criteria more efficient since the losses that belong to the detection are not attributed to the state which is characterized \cite{Checchinato2024}. Here we focus on a different aspect of lossy detection, namely, the ability to certify stellar ranks higher than one with a minimum  number of click detectors.}
Admittedly, the considered setup can certify high stellar ranks of a narrow subset of states only. Nevertheless, we believe that it is  of fundamental interest to investigate what are the minimum experimental resources required for certification of higher stellar ranks, and the setup considered in this work is remarkably simple.

The rest of the paper is organized as follows. In Sec. II we review some basic definitions and facts concerning the stellar rank and  its certification. We also specify and discuss a property of quantum measurement diagonal in Fock basis that  makes it useful for certification of higher stellar ranks. In Sec. III we describe the considered measurement scheme. In Sec. IV we present the numerically derived stellar rank thresholds for stellar rank witness based on the probability that one detector clicks and the other does not click, and we discuss  dependence of the thresholds on the setup parameters. In Section V we extend our analysis to more refined class of witnesses based on combination of probabilities of clicks of one detector and both detectors.  \tcr{Various experimental factors that can influence the ability to certify higher stellar ranks are discussed in Sec.~VI.} Finally, Sec. \tcr{VII} contains a brief summary and conclusions.

\section{Stellar rank}
A pure single-mode quantum state of light possesses stellar rank $m$ if it can be expressed as  \cite{Chabaud2020,Lachman2019}
\begin{equation}
|\psi\rangle= \hat{U}_G \sum_{n=0}^m c_n |n\rangle,
\label{psimpure}
\end{equation}
where $\hat{U}_G$ denotes an arbitrary Gaussan unitary operation, $|n\rangle$ stands for Fock states, and $c_{m}\neq 0$.

In practice the generated quantum states are affected by various imperfections, noise, and decoherence. It is therefore important to generalize the concept of stellar rank to arbitrary mixed states described by density matrix $\hat{\rho}$. A mixed state $\hat{\rho}$ has a stellar rank $m$ if it can be expressed as a statistical mixture of pure states with stellar rank at most $m$ but cannot be expressed as a mixture of pure states with stellar rank lower than $m$ only \cite{Chabaud2020,Chabaud2021}. More formally and rigorously, consider all possible pure-state decompositions $\hat{\rho}=\sum_{i} p_i |\psi_i\rangle\langle \psi_i|$, $p_i \geq 0$,  and denote by $m_i$ the stellar rank of $|\psi_i\rangle$.   The stellar rank $m_{\hat{\rho}}$ of mixed state $\hat{\rho}$ can be defined as \cite{Chabaud2021}
\begin{equation}
m_{\hat{\rho}}= \inf_{p_i,|\,\psi_i\rangle} \left(\sup_i m_i\right),
\end{equation}
and the infimum is calculated over all possible pure-state decompositions.

In experiments, we are interested in characterization and certification of stellar rank of the generated states. As already noted in the Introduction, various witnesses of stellar rank have been proposed and utilized in the literature. The simplest witnesses are based on fidelity $F$ of a given state with some pure state with stellar rank $m$ such as the Fock state $|m\rangle$ \cite{Chabaud2021,Podhora2022,Fadrny2024,Fadel2025}. More generally, one can consider a linear witness  specified  by some Hermitian operator $\hat{W}$ 
\cite{Chabaud2021,Fiurasek2022,Lachman2019} and determine stellar rank thresholds $W_m$ such that the stellar rank $m$  is certified if 
\begin{equation}
\mathrm{Tr} [\hat{W} \hat{\rho}] > W_m.
\end{equation}
\tcr{The stellar rank witnesses are conceptually similar to entanglement witnesses \cite{Guhne2009,Horodecki2009} since in both cases the goal is to prove that a given quantum state $\hat{\rho}$ does not belong to some convex set of states. In the present case these convex sets $\mathcal{S}_{m}$ are formed by all pure states (\ref{psimpure}) and their convex mixtures. Therefore, the  witness threshold $W_m$ can be determined by maximization of $\mathrm{Tr}[\hat{W}\hat{\rho}]$ over $\mathcal{S}_{m-1}$. Thanks to the linearity of the witness, it actually suffices to perform the maximization over all pure states (\ref{psimpure}) \cite{Chabaud2021,Fiurasek2022}.} 

\tcr{More explicitly, we can express $W_m$ as \cite{Chabaud2021,Fiurasek2022}}
\begin{equation}
\tcr{W_{m}=\sup_{|\phi_{m-1}\rangle, \,\hat{U}_{G}} \langle \phi_{m-1}| \hat{U}_G \hat{W} \hat{U}_G^\dagger |\phi_{m-1}\rangle,}
\label{Wmdefinition}
\end{equation}
where the supremum is taken over all unitary Gaussian operations $\hat{U}_G$ and over all finite superpositions of Fock states $|\phi_{m-1}\rangle=\sum_{n=0}^{m-1} c_n |n\rangle$.
Optimization over the pure states $|\phi_{m-1}\rangle$ leads to the maximum eigenvalue of the operator $\hat{U}_G \hat{W} \hat{U}_G^\dagger $ projected onto the subspace $\mathcal{H}_{m-1}$  spanned by the Fock states up to $m-1$. Let us define the corresponding projector
\begin{equation}
\hat{Z}_{m-1}= \sum_{n=0}^{m-1}|n\rangle\langle n|.
\end{equation}
We have \cite{Fiurasek2022,Miyata2016}
\begin{equation}
\tcr{W_{m}=\sup_{\hat{U}_G} \left\{\max \mathrm{eig} \left[ \hat{Z}_{m-1} \hat{U}_{G} \hat{W} \hat{U}_G^\dagger \hat{Z}_{m-1} \right] \right\}.}
\label{Wmsimplified}
\end{equation}
The remaining optimization over all single-mode Gaussian unitaries can be performed numerically, as discussed in more detail in Sec.~IV.

Consider now a measurement scheme described by POVM elements $\hat{\Pi}_{k}$ that are diagonal in Fock basis. Let $p_{k}(n)=\langle n|\hat{\Pi}_k|n\rangle$ denote the probability of measurement outcome $k$ for input Fock state $|n\rangle$. Let us consider a witness formed by a specific POVM element $\hat{W}=\hat{\Pi}_k$.  A sufficient condition for this witness  to certify the stellar rank $m$ of Fock state $|m\rangle$ is that 
\begin{equation}
p_{k}(m) > p_k(n), \qquad n \neq m.
\end{equation}
It is important that the inequality is strict and there exists a finite gap between $p_{k}(m)$ and the other probabilities,
\begin{equation}
\Delta=\inf_{n\neq m} [p_{k}(m)-p_{k}(n)] >0.
\end{equation}
Nonzero $\Delta$  ensures that the measured probability $p_k=\mathrm{Tr}[\hat{\Pi}_k\hat{\rho}]$ provides a lower bound on the fidelity $F_m$ with Fock state $|m\rangle$:
\begin{equation}
F_{m}=\langle m|\hat{\rho}|m\rangle \geq 1-\frac{p_k(m)-p_{k}}{\Delta}.
\label{Fmbound}
\end{equation}
\tcr{As shown in Refs.~\cite{Filip2011,Chabaud2021} fidelities with Fock state represent witnesses of quantum non-Gaussianity and stellar rank, associated with witness operator $\hat{W}=|m\rangle\langle m|$. In particular, for any Fock state $|m\rangle$ it is possible to determine a threshold $F_{m,\mathrm{th}}$ such that the state has a stellar rank $m$ if $F_m > F_{m,\mathrm{th}}$, and the fidelity thresholds are strictly smaller than $1$. For any state, $p_k\geq p_{k}(m)F_m$ holds. This, together with Eq. (\ref{Fmbound}), implies that the lower bound on fidelity $F_m$ will never be lower than 
\begin{equation}
F_{m,\mathrm{min}}= 1-\frac{p_k(m)}{\Delta}(1-F_m),
\end{equation}
where $F_m$ is the true value of the fidelity. Stellar rank $m$ will thus be certified for any state that satisfies $F_{m,\mathrm{min}}>F_{m,\mathrm{th}}$, which can be equivalently expressed as
\begin{equation}
F_{m}> 1-\frac{\Delta}{p_{k}(m)}(1-F_{m,\mathrm{th}}).
\label{Fboundconservative}
\end{equation}
Note that, depending on the exact shape of the photon number distribution of state $\hat{\rho}$, certification of  stellar rank $m$ may be possible even for fidelities $F_m$  smaller than the bound in Eq. (\ref{Fboundconservative}). }

\begin{figure}[t]
\centerline{\includegraphics[width=0.9\linewidth]{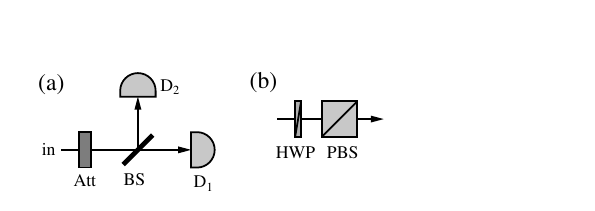}}
\caption{Proposed measurement scheme (a). The input signal is attenuated by attenuator (Att) and split at a beam splitter (BS) to be detected by two binary click detectors D$_1$ and D$_2$. Example of tunable attenuator for input linearly polarized light (b). The attenuator is formed by a sequence of half-wave plate (HWP) and polarizing beam splitter (PBS).}
\label{figsetup}
\end{figure}

\section{Measurement scheme} 

After these general considerations, let us focus on the experimental scheme depicted in Fig.~\ref{figsetup}. The input signal is first attenuated by an attenuator with intensity transmittance $\eta_A$ and then it is split into two parts at a beam splitter (BS) with transmittance $T$ and reflectance $1-T$. The transmitted and reflected modes are detected with binary click detectors D$_1$ and D$_2$ that distinguish the presence and absence of photons. 
The detection efficiency $\eta_D$ of the detectors can be incorporated into the overall loss factor $\eta=\eta_A\eta_D $.  The POVM element associated with a click of the binary  detector with total detection efficiency $\eta$ reads as
\begin{equation}
\hat{\Pi}_{C}=\sum_{n=1}^\infty [1-(1-\eta)^n]|n\rangle\langle n|.
\end{equation}
We note that the Hanbury Brown-Twiss setup depicted in Fig.~\ref{figsetup} has been utilized in the first experimental test of quantum non-Gaussianity of approximate single-photon states \cite{Jezek2011} and analytical criteria of quantum non-Gaussianity were derived for this scheme in Refs. \cite{Lachman2013,Fiurasek2021}.

\begin{figure}[t]
\centerline{\includegraphics[width=\linewidth]{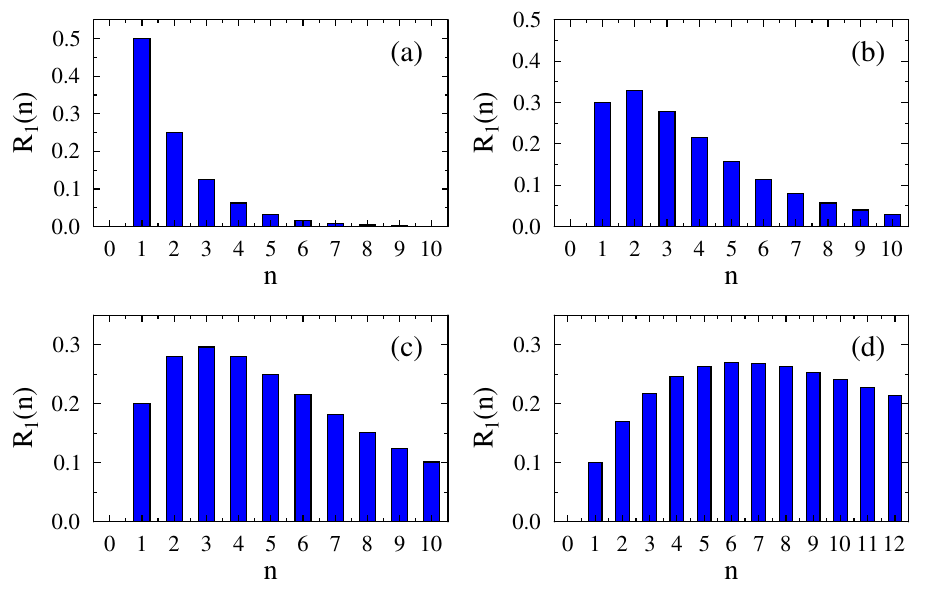}}
\caption{Probabilities $R_1(n)$ of click of detector D$_1$ and simultaneous no-click of D$_2$ for input Fock states $|n\rangle$ are plotted for balanced detection scheme with $T=0.5$ and four different total detection efficiencies 
$\eta=1$ (a), $\eta=0.6$ (b), $\eta=0.4$ (c),  and $\eta=0.2$ (d).}
\label{figReta}
\end{figure}

The probability $R_1$ that only detector D$_1$ in Fig.~\ref{figsetup}(a) clicks can be expressed as a difference of probability that D$_2$ does not click irrespective of the response of D$_1$, and the probability that both D$_1$ and D$_2$ do not click. 
For input Fock state $|n\rangle$ we obtain
\begin{equation}
R_1(n)=[1-\eta (1-T)]^n-(1-\eta)^n,
\end{equation}
and for $\eta=1$ this expression simplifies to 
\begin{equation}
R_1(n)=T^n-\delta_{n,0}.
\end{equation}
For $\eta=1$, probability $R_1(n)$ is maximized at $n=1$ and then monotonically decreases with increasing $n$. This is illustrated in Fig.~\ref{figReta}(a) for balanced setup with $T=\frac{1}{2}$. However, with decreasing $\eta$ the maximum of $R_1(n)$ shifts to higher photon numbers $n$, as also illustrated in Fig.~\ref{figReta}. 

In order to get some insight into the mechanism which shifts the maximum detection probabilities to higher  Fock states it is instructive to compare results for Fock states $|1\rangle$  and $|2\rangle$. To simplify the discussion it is convenient to attribute all losses to the attenuator Att. For input Fock state $|1\rangle$ the probability that D$_1$ clicks reads as $R_1(1)=\eta T $, which is the product of probabilities that a single photon passes through the attenuator and reaches the correct output port with detector D$_1$. For the input Fock state $|2\rangle$   both photons will pass through the attenuator with probability $\eta^2$ and with probability $2\eta(1-\eta)$ only one of the photons passes through the attenuator. Consequently, the probability $R_1(2)$ can be expressed as a sum of two contributions,
\begin{equation}
R_1(2)=\eta^2 T^2+2\eta(1-\eta)T.
\end{equation}
Observe that for $\eta<\frac{1}{2}$ the attenuated two-photon Fock state contains higher single-photon contribution than the attenuated single-photon state since $2\eta(1-\eta)>\eta$.
More generally, by comparinng the single-click probabilities $R_1(1)$ and $R_1(2)$  we find that  the peak of the function $R_1(n)$ will be at $n \geq 2$ when
\begin{equation}
\eta < \frac{1}{2-T}.
\end{equation}

\begin{figure}[t]
\centerline{\includegraphics[width=0.9\linewidth]{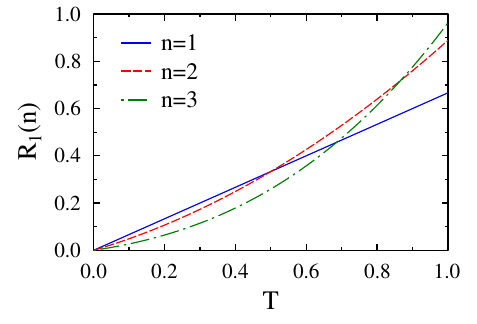}}
\caption{Probability $R_1(n)$ of click of detector D$_1$ and simultaneous no-click of D$_2$ for input Fock state $|n\rangle$ is plotted as function of $T$ for $\eta=\frac{2}{3}$ and $n=1$ (blue solid line), $n=2$ (red dashed line), and $n=3$ (green dotted-dashed line).}
\label{figRTdependence}
\end{figure}

The dependence of $R_1(n)$ on $T$ is plotted in Fig.~\ref{figRTdependence} for fixed $\eta=\frac{2}{3}$, which illustrates the role of the beam-splitter transmittance. We can see that as $T$ increases the peak of $R_1(n)$ shifts to higher photon numbers $n$.

\tcr{In order to obtain additional insight into the dependence of the shape of $R_1(n)$ on $\eta$ and $T$ we shall treat for a while $n$ as continuous parameter and seek $n$ that maximizes $R_1(n)$ by solving the extremal equation $\frac{d R_1(n)}{d n}=0$. This yields
\begin{equation}
n_{\mathrm{max}}= \frac{\ln\displaystyle{\frac{ \ln [1-\eta(1-T)]}{\ln(1-\eta)}}}{\ln(1-\eta)-\ln [1-\eta(1-T)]}.
\label{nmaxanalytical}
\end{equation}
In practice, we should consider the $n_{\mathrm{max}}$ rounded up and down to find the integer number of photons which maximizes $R_1(n)$. Formula (\ref{nmaxanalytical}) simplifies in the limit of $\eta \ll 1$,
\begin{equation}
n_{\mathrm{max}} \approx \frac{\ln(1-T)^{-1}}{\eta T}.
\label{nmaxasymptotic}
\end{equation}
Formulas (\ref{nmaxanalytical}) and (\ref{nmaxasymptotic}) approximately quantify the maximum stellar rank that can be certified based on the knowledge of the probability $R_1$.
In accordance with the results plotted in Figs. 2 and 3 these formulas indicate that the maximum certifiable stellar rank increases with increasing losses and also increases with increasing $T$.
}

\tcr{
The mechanism by which the losses help to increase the certifiable stellar rank is not limited to the specific detection scheme in Fig.~1 but is rather general. To illustrate this, we consider now a projection on Fock state $|k\rangle$ preceded by losses. The probability of detection of $k$ photons for input $n$-photon Fock state $|n\rangle$ reads as
\begin{equation}
P_k(n)={ n\choose k} \eta^k (1-\eta)^{n-k}, \qquad n\geq k,
\end{equation}
and the probability vanishes for $n<k$.  Specifically for $k=1$ we have 
\begin{equation}
P_{1}(n)=n\eta (1-\eta)^{n-1},
\end{equation}
and the maximum of $P_{1}(n)$ can be found analytically. After some algebra we get
\begin{equation}
n_{\mathrm{max}}= -\frac{1}{\ln(1-\eta)}.
\end{equation}
In the limit of low $\eta$ we obtain $n_{\mathrm{max}}\approx 1/\eta$. For  $k>1$  a similar approximate asymptotic formula can be derived,  $n_{\mathrm{max}}\approx k/\eta$. The scaling $n_{\mathrm{max}} \propto1/\eta$ thus emerges as  rather universal.
}

\section{Stellar rank thresholds}

In order to determine the stellar rank thresholds (\ref{Wmdefinition}) we first write the POVM element corresponding to click of D$_1$ and no-click of D$_2$:
\begin{equation}
\hat{W}=\hat{R}_1=\sum_{n=0}^\infty R_1(n) |n\rangle\langle n| .
\label{Wdefinition}
\end{equation}
This explicitly reads as
\begin{equation}
\hat{W}=\sum_{n=0}^\infty [1-\eta(1-T)]^n|n\rangle\langle n| - \sum_{n=0}^\infty (1-\eta)^n |n\rangle\langle n|. 
\label{Wexplicit}
\end{equation}
To proceed, we recall the expression for density matrix of a thermal state with mean number of photons $\bar{n}$:
\begin{equation}
\hat{\rho}_{\mathrm{th}}(\bar{n})=\sum_{n=0}^\infty \frac{\bar{n}^n}{(\bar{n}+1)^{n+1}}|n\rangle\langle n|.
\label{rhothermal}
\end{equation}
By comparing Eqs. (\ref{rhothermal}) and (\ref{Wexplicit}) we find that  the formula (\ref{Wexplicit}) can be interpreted as a weighted linear combination of density matrices of two thermal states,
\begin{equation}
\hat{W}=(1+\bar{n}_1) \rho_{\mathrm{th}}(\bar{n}_1)- (1+\bar{n}_2) \rho_{\mathrm{th}}(\bar{n}_2),
\label{Wthermal}
\end{equation}
where
\begin{equation}
\bar{n}_1=\frac{1}{\eta (1-T)}-1,  \qquad \bar{n}_2=\frac{1}{\eta}-1.
\label{nbars}
\end{equation}
To explicitly perform the maximization in Eq. (\ref{Wmsimplified}), we need to calculate density matrix elements in Fock basis of mixed single-mode Gaussian states
\begin{equation}
\hat{\rho}_G=\hat{U}_G \hat{\rho}_{\mathrm{th}} (\bar{n}) \hat{U}_G^\dagger,
\label{rhoGdefinition}
\end{equation}
where $\hat{U}_G$ is an arbitrary Gaussian operation.

According to the Bloch-Messiah decomposition we can express any single-mode unitary  Gaussian operation as a sequence of phase shifts, single-mode suqeezing, and coherent displacement \cite{Braunstein2005PRA},
\begin{equation}
\hat{U}_{G}= e^{i \hat{n}\phi} \hat{D}(\alpha) \hat{S}(r) e^{i\hat{n}\theta},
\label{UGfactorization}
\end{equation}
where
\begin{equation}
\hat{D}(\alpha)=e^{\alpha \hat{a}^\dagger - \alpha^\ast \hat{\alpha}}, \qquad \hat{S}(r)=e^{\frac{r}{2}(\hat{a}^{\dagger 2}-\hat{a}^2)}.
\label{DSdefinition}
\end{equation}
The phase shift $\theta$ is irrelevant because the thermal state is diagonal in Fock basis. Similarly, the second phase shift $\phi$ can be omitted because $e^{i\phi \hat{n}}$ commutes with the projector $\hat{Z}_{m-1}$ and leaves the subspace $\mathcal{H}_{m-1}$ invariant. We can therefore set $\theta=\phi=0$ without any loss of generality and consider $\hat{U}_G=\hat{D}(\alpha)\hat{S}(r).$

Analytical formulas for density matrix elements  $\rho_{G,mn}=\langle m|\hat{\rho}_{G}| n\rangle$  of the displaced squeezed thermal state have been derived in Refs. \cite{Vourdas1986,Marian1993}.  For completeness, we recall that these density matrix elements can be calculated  using the well-known fact that the Husimi $Q$ function
\begin{equation}
Q(\beta,\beta^\ast)=\frac{1}{\pi}\langle \beta|\hat{\rho}|\beta\rangle
 \end{equation}
  is a generating function of $\rho_{mn}$,
  \begin{equation}
  \rho_{mn}= \frac{\pi}{\sqrt{m!n!}} \frac{\partial ^{m+n}}{\partial \beta^{\ast m} \partial \beta^{ n}}\left.\left[Q(\beta,\beta^\ast) e^{|\beta|^2}\right ] \right|_{\beta=\beta^\ast=0}.
  \label{rhomngenerating}
  \end{equation}
Here $\beta$ and $\beta^\ast$  are formally treated as two independent variables. The displaced squeezed thermal state $\hat{\rho}_G$ possesses Gaussian Husimi $Q$ function
 \begin{equation}
 Q(\beta,\beta^\ast)=\frac{1}{\pi}\frac{1}{\sqrt{V_R V_I}}\exp\left[-\frac{(\beta_R-\alpha_R)^2}{V_R}- \frac{(\beta_I-\alpha_I)^2}{V_I}\right],
 \label{HusimiG}
 \end{equation}
where 
 \begin{equation}
V_R=\frac{1}{2}[(2\bar{n}+1)e^{2r}+1], \qquad V_I=\frac{1}{2}[(2\bar{n}+1)e^{-2r}+1],
\end{equation}
and the complex amplitudes are decomposed into their real and imaginary parts, $\alpha=\alpha_R+i\alpha_I$ and  $\beta=\beta_R+i\beta_I$.
  On inserting the expression (\ref{HusimiG}) into formula (\ref{rhomngenerating})  one obtains \cite{Marian1993}
\begin{eqnarray}
\rho_{G,mn}&=& \sqrt{\frac{m! n!}{V_R V_I}}\exp\left(-\frac{\alpha_R^2}{V_R}- \frac{\alpha_I^2}{V_I}\right) \nonumber \\
& &\times \sum_{j=0}^{\min(m,n)} \frac{A^{m+n-2j }B^j}{j!(m-j)!(n-j)!} H_{m-j}(\omega) H_{n-j}(\omega^\ast). \nonumber \\
\end{eqnarray}
Here $ H_n(\cdot)$  denotes Hermite polynomial of degree $n$, and the parameters $A$, $B$, and $\omega$ read as
\begin{eqnarray}
&\displaystyle A=\sqrt{\frac{1}{4V_R}-\frac{1}{4V_{I}}}, \qquad B=1-\frac{1}{2V_R}-\frac{1}{2V_I},& \nonumber \\[2mm]
&\displaystyle \omega= \frac{1}{A}\left(\frac{\alpha_{R}}{2V_R}+\frac{i\alpha_I}{2V_I}\right).& 
\end{eqnarray}

The optimization of the stellar rank thresholds $W_m$ over three real parameters $\alpha_R$, $\alpha_I$, and $r$ was performed numerically by an exhaustive search. The numerics reveals that it is optimal to displace the state along one of the principal squeezing axes, which is consistent with previous observations \cite{Filip2011,Lachman2013,Lachman2019}. The final optimization was therefore performed over real non-negative $\alpha$ and real $r$ (both positive and negative). Numerical results for a balanced scheme with $T=\frac{1}{2}$ are summarized in Table I. In addition to the stellar rank thresholds we also specify in the table the maximum possible value of the witness, i.e., maximum of $R_{1}(n)$ over all $n$ for given $\eta$. We can see that the maximum stellar rank that can be certified by measurement of the single-click probability $R_1$ increases with decreasing detection efficiency $\eta$, which is consistent with the shift of the maximum of $R_1(n)$ to higher $n$ [cf. Fig.~\ref{figReta} and Eq. (\ref{nmaxasymptotic})]. However,  \tcr{ the interval $W_{\mathrm{max}}-W_m$  becomes very narrow as $m$ increases}. This indicates that the witness can certify only a narrow subset of states with stellar rank $m$ and the measurement precision would have to be very high to
achieve reliable certification. In Table II we  present the stellar rank thresholds for several different values of beam-splitter transmittance $T$. In accordance with the behavior observed 
in Fig.~3, then maximum certifiable stellar rank increases with increasing $T$.

\begin{table}[t]
\begin{ruledtabular}
\begin{tabular}{ccccc} 
 $\eta$ &  1 & 0.6 & 0.4 & 0.2 \\ \hline
$W_1$ & 0.2979  & 0.2784 & 0.2696 & 0.2606 \\
$W_2$ & --- & 0.3119 & 0.2810 & 0.2638 \\
$W_3$ & ---  & --- & 0.2881 & 0.2654 \\
$W_4$ & ---  & --- &  --- &  0.2664 \\
$W_5$ & ---  & --- &  --- & 0.2672 \\ 
$W_6$ & ---  & --- &  --- & 0.2679 \\  \hline 
$W_{\mathrm{max}}$ & 0.5000  & 0.3300 & 0.2960 & 0.2693 \\  
\tcr{$n_{\mathrm{max}}$}  & \tcr{1} & \tcr{2} & \tcr{3} & \tcr{6} 
\end{tabular}
\end{ruledtabular}
\caption{Stellar rank thresholds $W_m$ for the witness $\hat{W}=\hat{R}_1$ specified in Eq. (\ref{Wdefinition}). The thresholds $W_m$ are presented for $T=0.5$ and four different values of the total detection efficiency $\eta$. The witnesses $W_m$ are rounded up to four decimal places. \tcr{The last two rows specifiy the maximum physically achievable value of the witness and the Fock number $n_{\mathrm{max}}$ for which $R_1(n_{\mathrm{max}})=W_{\mathrm{max}}$ is reached.}} 
\end{table}

\begin{table}[t]
\begin{ruledtabular}
\begin{tabular}{ccccc} 
 $T$ &  0.4 & 0.6 & 0.8 & 0.9 \\ \hline
$W_1$ &  0.2055 & 0.3535 & 0.5660 & 0.7234 \\
$W_2$ & 0.2235  & 0.3732 & 0.5822 & 0.7345 \\
$W_3$ & ---  & ---  &  0.5914 &  0.7403  \\
$W_4$  & ---  & ---  & ---  & 0.7449 \\  \hline 
$W_{\mathrm{max}}$ & 0.2400 & 0.3900 & 0.6040 & 0.7520 \\
\tcr{$n_{\mathrm{max}}$}  & \tcr{2} & \tcr{2} & \tcr{3} & \tcr{4} 
\end{tabular}
\end{ruledtabular}
\caption{The same as Table I but the total detection efficiency is fixed at $\eta=0.5$ and the stellar rank thresholds are presented for four different values of beam splitter transmittance $T$. }
\end{table}

\section{Witnesses based on $R_1$ and $ R_2$}

\begin{figure}[t]
\includegraphics[width=\linewidth]{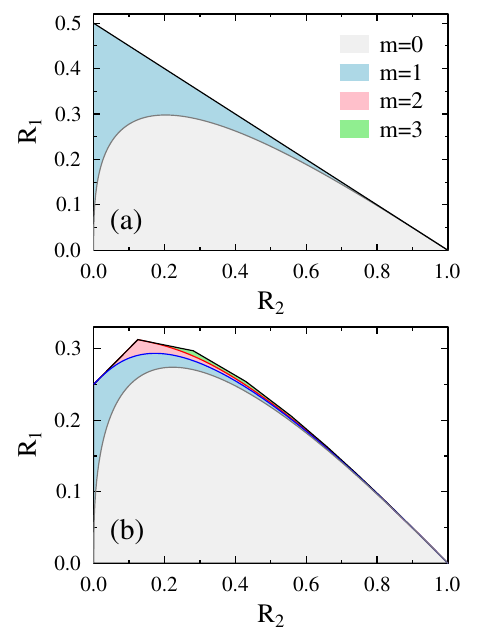}
\caption{Witnesses of stellar rank based on the pair of probabilities $R_1$ and $R_2$ are plotted for balanced setup with $T=0.5$, and total detection efficiency $\eta=1$ (a)  and $\eta=0.5$ (b). The color of each area indicates the minimum stellar rank that is certified when the point $[R_2,R_1]$ lies in that area. Points in the light gray area are compatible with Gaussian states and their mixtures. For $\eta=1$, only stellar rank $1$ can be certified (the light blue area), while stellar ranks higher than one become certifiable for $\eta=0.5$. The black lines indicate the boundary of all physically achievable probability pairs $[R_2,R_1]$. See text for additional details. }
\label{figW}
\end{figure}

In this section we consider more general class of stellar rank witnesses based on the pair of probabilities of $R_1$ and $R_2$, where $R_2$ denotes the probability that both detectors D$_1$ and  D$_2$ click simultaneously \cite{Jezek2011,Straka2018,Lachman2019,Checchinato2024}.  
We shall first consider balanced detection scheme with $T=\frac{1}{2}$. The probability $R_2(n)$ for input Fock state $n$ is then given by
\begin{equation}
R_2(n)=1-2\left(1-\frac{\eta}{2}\right)^n+(1-\eta)^n.
\end{equation}
Here we made use of the identity $R_2=1-2R_1-R_0$ valid for balanced two-detector scheme, and $R_0(n)=(1-\eta)^n$ is the probability that none of the detectors click \tcr{for input Fock state $|n\rangle$}.
The POVM element $\hat{R}_2$ associated with this detector outcome reads as
\begin{equation}
\hat{R}_2 =\sum_{n=0}^\infty R_2(n)|n\rangle\langle n|.
\end{equation}
Note that $\hat{R}_2 $ can  be expressed as a linear combination of identity operator and density matrices of two thermal states.

Following the long-established approach of Refs. \cite{Filip2011,Jezek2011,Lachman2019,Straka2018}, we can consider witnesses formed by linear combinations of $\hat{R}_1$ and $\hat{R}_2$,
\begin{equation}
\hat{W}_\lambda= \hat{R}_1 -\lambda \hat{R}_2.
\end{equation}
Explicitly, we get
\begin{equation}
\hat{W}_\lambda=(1+2\lambda)(1+\bar{n}_1) \rho_{\mathrm{th}}(\bar{n}_1)-(1+\lambda)(1+\bar{n}_2) \rho_{\mathrm{th}}(\bar{n}_2) -\hat{I}\lambda.
\end{equation}
Here $\hat{I}$ denotes the identity operator, $\bar{n}_1=2/\eta-1$ and $\bar{n}_2=1/\eta-1$.

The set of probability pairs $[R_2,R_1]$ that are achievable by quantum states with stellar rank at most $m-1$ forms a convex set $\mathcal{S}_{m-1}$ in a two-dimensional space, see Fig.~\ref{figW}. Determination of the stellar rank threshold $W_{\lambda,m}$ for the witness $\hat{W}_\lambda$  identifies a line that is tangent to the boundary of this convex set at some point. By varying the parameter $\lambda$ we can probe and find the whole boundary of this convex set $\mathcal{S}_{m-1}$.

For reference, we first plot in Fig.~\ref{figW}(a)  the results for $\eta=1$ \cite{Lachman2019}. In this case, the setup allows to certify only the stellar rank $1$. Note that analytical description of the quantum non-Gaussianity boundary depicted in Fig.~4(a) was derived in Ref. \cite{Lachman2013}. Aside from the boundary between the various sets $\mathcal{S}_{m}$ it is also important to identify the ultimate boundary of all physically achievable probability pairs $[R_2, R_1] $. For $\eta=1$, this boundary is formed by a triangle \cite{Lachman2013}, as illustrated in Fig.~\ref{figW}(a). The vertices of this triangle correspond to the vacuum state $[0,0]$, single-photon Fock state $[0, \frac{1}{2}]$, and a Gaussian quantum state with asymptotically infinite energy so that both detectors always click, $[1,0]$. Note that the probability pairs $[R_2(n),R_1(n)]$ for input Fock states with $n>1$ all lie on a straight line that connects the points $[0,\frac{1}{2}]$ and $[1,0]$. To show this, note that for  $\eta=1$ we get $R_2(n)=1-2/2^n+\delta_{n0}$ and $R_{1}(n)=1/2^n-\delta_{n0}$. For $n>0$ we thus obtain a simple linear relationship 
\begin{equation}
R_1(n)=\frac{1-R_2(n)}{2}.
\end{equation}

As illustrated in Fig.~\ref{figW}(b), this picture changes for $\eta<1$ and  stellar ranks higher than $1$ become certifiable for certain states. Note that the witnesses based on pair of probabilities $R_1$ and $R_2$ are strictly stronger than witness based just on $R_1$, because the knowledge of probability of simultaneous clicks of both detectors $R_2$  provides additional information  that imposes stricter limit on the maximum achievable probability $R_1$ of click of the first detector only. \tcr{ Note also that the quantum non-Gaussianity criterion for this type of lossy detection with a pair of click detectors was presented in Ref. \cite{Checchinato2024} [cf. the boundary of the gray area in Fig.~4(b)].}

For $\eta<1$ the boundary of the set of all physically achievable probability pairs $[R_2, R_1]$ is not a simple triangle but is instead formed by an infinite number of straight lines that connect the points $[R_2(n),R_1(n)]$ and $R_2(n+1),R_1(n+1)]$. The probability pairs $[R_2(n),R_1(n)]$ corresponding to various Fock states $|n\rangle$  do not lie on a single straight line anymore but on a concave curve (see Appendix for more details). The boundary of the convex hull of this set is thus formed by the above specified sequence of straight lines. This structure of the physical boundary suggests that stellar rank $m$ of arbitary high Fock state $|m\rangle$ is in principle certifiable by this scheme. However, for large $m$  the boundaries of the various sets $\mathcal{S}_m$ practically coincide.  Successful application of the witness to high Fock states may therefore require extremely high precision of the measurement of click probabilities $R_1$  and $R_2$.

\begin{figure}[t]
\includegraphics[width=\linewidth]{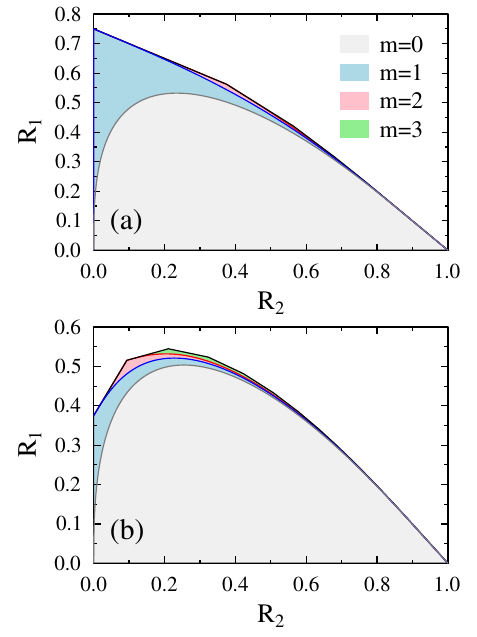}
\caption{The same as Fig.~\ref{figW}, but $T=0.75$. Witnesses of stellar rank based on the pair of probabilities $R_1$ and $R_2$ are plotted for total detection efficiency $\eta=1$ (a)  and $\eta=0.5$ (b). The color of each area indicates the minimum stellar rank that is certified when the point $[R_2,R_1]$ lies in that area. The black lines indicate the boundary of all physically achievable probability pairs $[R_2,R_1]$. }
\label{figWunbalanced}
\end{figure}

Finally, let us investigate also an unbalanced detection scheme with $T \neq \frac{1}{2}$. In this case, the degeneracy is lifted and probabilities of clicks of single detectors D$_1$ and D$_2$ differ.
The probability of simultaneous click of both detectors for input Fock state $|n\rangle $ can be expressed as 
\begin{equation}
R_2(n)=1-(1-\eta T)^n-[1-\eta(1-T)]^n+(1-\eta)^n,
\end{equation}
and the corresponding POVM element $\hat{R}_2$  can be written as a linear combination of identity operator and density matrices of three thermal states,
\begin{equation}
\hat{R}_2=\hat{I}-(1+\bar{n}_1) \rho_{\mathrm{th}}(\bar{n}_1)-(1+\bar{n}_3) \rho_{\mathrm{th}}(\bar{n}_3)+(1+\bar{n}_2) \rho_{\mathrm{th}}(\bar{n}_2),
\end{equation}
where $\bar{n}_1=1/[\eta(1-T)]-1$, $\bar{n}_2=1/\eta-1$, and $\bar{n}_3=1/(\eta T)-1$ [see also Eq. (\ref{nbars})]. 

In Figure~\ref{figWunbalanced} we plot the numerical results for $T=0.75$. We can see that for unbalanced detection scheme the witness based on the pair of probabilities of click of the first detector only $R_1$ and probability of simultaneous click click of both detectors $R_2$ can detect stellar ranks higher than $1$ even for $\eta=1$. However, the area corresponding to detection of higher stellar ranks is very small. When the detection efficiency is decreased to $\eta=0.5$  then also the area corresponding to the certification of stellar rank $3$ becomes  visible [see Fig.~\ref{figWunbalanced}(b)].

\tcr{\section{Experimental considerations}}

\tcr{Several experimental factors can influence the ability to certify stellar ranks higher than one with the proposed approach. First of all, the total number of measurements $N$ must be high enough to suppress the statistical errors. To simplify the analysis we shall consider here the stellar rank witnesses based solely on the probability $R_1=\sum _n d_n R_1(n)$, where $d_n=\langle n|\hat{\rho}|n\rangle$. The required number of measurements to certify stellar rank $m$  can be estimated by requiring that the gap $R_1-W_{m}$ between $R_1$ and the stellar rank threshold $W_m$ is larger than three standard deviations of the estimate of $R_1$:
\begin{equation}
R_1-W_m\gtrsim 3 \sqrt{ \frac{ R_1(1-R_1)}{N}} .
\label{samplingerror}
\end{equation}
 Formula (\ref{samplingerror})  yields
\begin{equation}
N \gtrsim   \frac{9R_1(1-R_1)}{(R_1-W_m)^2}.
\label{Nthreshold}
\end{equation}
Assuming that the actual value of $R_1$ is $W_{\mathrm{max}}$ (hence the characterized state is a pure Fock state), and 
plugging in the data from Tables I and II we find that reliable certification of stellar rank 2 is possible with  $N\approx 10^4$ samples, while certification of stellar rank $3$ may be possible with slightly less than  $10^5$ samples. The number of required samples $N$ will further increase for imperfect Fock states and is mainly sensitive to the width of the gap $R_1-W_m$.}

\tcr{
Dark counts represent another factor that can influence the measurement. We consider here a simple model of dark counts that appear with probability $P_D$ that does not depend on the state which is measured by the detector. Presence of dark counts modifies the expression for $R_1(n)$ as
\begin{equation}
R_1(n)=(1-P_D)[1-\eta(1-T)]^n-(1-P_D)^2(1-\eta)^n,
\end{equation}
and similar generalization can be performed also for $R_2(n)$. The stellar rank witnesses and thresholds  can be determined for this modified and extended detection model in the same way as for the original model without dark counts. In practice, the dark counts can be of the order of 100 per second or even less. Considering measurement with detection window whose width  is several nanoseconds, the probability of dark counts will be $P_D  \lesssim10^{-6}$, hence typically negligible. }

\tcr{Finally, the precision of calibration of the parameters $\eta$  and $T$ can also play important role. It may be particularly simple to achieve the balanced operation $T=\frac{1}{2}$, as in this setting the response of the two detectors in Fig.~1 shall be the same (up to statistical fluctuations). Precise absolute calibration of losses may be more challenging. Let us suppose that $\eta$ can be calibrated with mean value $\bar{\eta}$ and residual uncertainty $\Delta \eta$. Then one should consider the stellar rank threshold $W_m(\eta)$ as a function of $\eta$ and choose the maximum in the interval of possible values of $\eta$,
\begin{equation}
W_m=\max_{\eta\in \mathcal{I}} W_{m}(\eta), \qquad \mathcal{I}=[\bar{\eta}-\Delta \eta,\bar{\eta}+\Delta \eta].
\end{equation}
As can be seen from data in Tables I and II, the stellar rank thresholds typically increase with increasing $\eta$. A conservative choice will therefore usually be to consider the maximum possible value of $\eta$, given by $\bar{\eta}+\Delta\eta$. This choice will conservatively assign part of the actual losses to the incoming state. Increased stellar rank thresholds will lead to increased numbers of required samples $N$, as indicated by Eq. (\ref{Nthreshold}). Moreover, if the uncertainty of $\eta$ will be large enough, then the certification of stellar ranks higher than one may become unfeasible. Numerical calculations for $T=0.5$ and $\eta=0.6$ show that the threshold $W_{2}=0.3119$ increases to $W_{2}=0.3144$ for $\Delta \eta=0.01$ and becomes $W_2=0.3265$ for $\Delta \eta=0.05$, which is already very close to the maximum possible value $W_{\mathrm{max}}=0.3300$. This example illustrates that the precision of calibration of $\eta$ should be of the order of $0.01$.}

\tcr{
As a final remark, we note that the spatial multiplexing utilized in Fig.~1 can be replaced by temporal multiplexing. This alternative approach requires only a single detector that is preceded by tunable losses \cite{Mogilevtsev1998,Rossi2004}.  The probabilities $R_1$ and $R_2$ can be estimated from the measured  probabilities of non-clicks of the detector for three levels of transmittances of the lossy channel: $\eta$, $\eta T$, and $\eta (1-T)$. For balanced setup with $T=\frac{1}{2}$ this reduces to just two different loss factors  $\eta$ and $\eta/2$.}

\section{Conclusions}

In conclusion, we have shown that \tcr{stellar ranks higher than $1$}  can be certified with a very simple experimental setup that consists of a beam splitter, two click detectors, and an attenuator. \tcr{The price to pay for the experimental simplicity is that the successful certification of higher stellar ranks  is only possible for a limited set of states, in particular states that are sufficiently close to ideal Fock states.}
We have explained how the reduced total detection efficiency helps to certify stellar rank of higher Fock states, and we have numerically calculated the stellar rank thresholds for the considered stellar rank witnesses. For the simple witness based on probability that D$_1$ clicks and D$_2$ does not click we have observed that the maximum detectable stellar rank increases with decreasing detection efficiency and increases with increasing beam-splitter transmittance.  The ability to certify higher stellar ranks can be further improved by considering more refined stellar rank witnesses that combine probability $R_1$ and probability $R_2$ that both detectors click simultaneously. In this latter case stellar ranks higher than $1$ can be certified even with perfect detectors and $\eta=1$ by utilizing an unbalanced scheme with $T\neq 0.5$. Nevertheless, even in such case losses can be helpful. 
Our results provide unique insights into the capabilities of the multiplexed click detectors and reveal that high stellar ranks can be certified even with a very simple scheme that involves only two detectors.

\begin{acknowledgments}
The author acknowledges support by the project OP JAC, Reg. No. CZ.02.01.01/00/23\_021/0008790,  of the Ministry of Education, Youth, and Sports of the Czech Republic and EU.

\end{acknowledgments}

\section*{Data availability statement}
The data that support the findings of this article are openly available \cite{CertificationZenodo2025}.

\appendix*

\section*{Appendix}
Here we discuss the properties of a parametric curve specified by $R_2(n)$ and $R_1(n)$, where $n$ is now treated as a continuous parameter and $T=\frac{1}{2}$. Recall that 
\begin{equation}
R_1(n)=\left(1-\frac{\eta}{2}\right)^n-(1-\eta)^n,
\end{equation}
and
\begin{equation}
R_2(n)=1-2\left(1-\frac{\eta}{2}\right)^n+(1-\eta)^n.
\end{equation}
This parametric curve specifies $R_1$ as function of $R_2$. We want to prove that this function is concave for $n\geq 1$. First observe that $R_2(n)$ is monotonically increasing function of $n$ for $n \geq 1$. Clearly, adding more photons to the input can only increase the probability that both detectors click simultaneously. More formally, since $n$ is now treated as a continuous parameter, we evaluate the derivative
\begin{widetext}
\begin{equation}
\frac{d R_2(n)}{d n}=\left(1-\frac{\eta}{2}\right)^n \left[\left(\frac{2-2\eta}{2-\eta}\right)^n \ln(1-\eta)-2\ln\left(1-\frac{\eta}{2}\right)\right].
\end{equation}
We want to prove that this function is positive for $n\geq 1$. Since $\ln(1-\eta)<0$ and $ (2-2\eta)/(2-\eta)<1$ it suffices to prove positivity for $n=1$. This leads us to the expression
\begin{equation}
(1-\eta)\ln(1-\eta)-2 \left(1-\frac{\eta}{2}\right)\ln\left(1-\frac{\eta}{2}\right)
\end{equation}
which is positive for $0<\eta<1$.
Since $R_2(n)$ monotonically increases with $n$, in order to prove the concavity it suffices to consider the signed curvature of the curve, defined as
\begin{equation}
k= \frac{R_2^\prime R_1^{\prime\prime}-R_1^\prime R_2^{\prime\prime}}{(R_1^{\prime^2} +R_2^{\prime 2})^{3/2}}.
\end{equation}
Here the prime denotes derivative with respect to $n$. Since we are only interested in the sign of $k$, it suffices to consider the function
\begin{equation}
f(n)=R_2^\prime R_1^{\prime\prime}-R_1^\prime R_2^{\prime\prime}.
\end{equation}
After some algebra, we get
\begin{equation}
f(n)=(1-\eta)^n \left(1-\frac{\eta}{2}\right)^n\ln(1-\eta)\ln\left(1-\frac{\eta}{2}\right) \ln \frac{2-2\eta}{2-\eta}.
\end{equation}
For $0<\eta<1$ this function is negative for all $n \geq 1$. This proves that the points $[R_2(n),R_1(n)]$ lie on a concave curve when $T=\frac{1}{2}$.
\end{widetext}

\end{document}